\documentclass[showpacs,preprintnumbers,amsmath,amssymb,endfloats*]{revtex4}
\usepackage{graphicx}

\begin{document}

\thispagestyle{empty}
\title{Demonstration of the difference Casimir force 
for samples with different charge carrier densities
}

\author{F.~Chen${}^{1}$, G.~L.~Klimchitskaya${}^2$,
V.~M.~Mostepanenko${}^3$, and U.~Mohideen${}^1$}

\affiliation{${}^{1}$Department of Physics, 
University of California,
Riverside, California 92521, USA. \\
${}^2$North-West Technical 
University, Millionnaya St. 5, St.Petersburg, 191065, Russia.\\
${}^3$Noncommercial Partnership ``Scientific Instruments'', 
Tverskaya St.{\ }11, Moscow, 103905, Russia.}

\begin{abstract}
A measurement of the Casimir force between a gold coated sphere and
two Si plates of different carrier densities is performed
using a high vacuum based atomic force microscope. 
The results are compared with the Lifshitz theory and 
good agreement is found. Our experiment demonstrates that by
changing the carrier density of the semiconductor plate
by several orders of magnitude it is
possible to modify the Casimir interaction. This result may find
applications in nanotechnology.
\end{abstract}

\pacs{12.20.Fv, 12.20.Ds, 68.37.Ps, 73.25.+i}

\maketitle

The Casimir effect \cite{1} implies that there is a force acting
between closely spaced neutral bodies determined by the zero-point
oscillations of the electromagnetic field. 
In the last few years the Casimir force was extensively
investigated experimentally (see, e.g., Refs,~\cite{2,3,4,5,6,7,8}
and review \cite{9}). It has found many diverse applications
ranging from Bose-Einstein condensation \cite{10}, carbon nanotubes
\cite{11}, and to the testing of predictions of new physics beyond
the standard model \cite{7,8,9}. One of the most important
applications of the Casimir effect is in the design, fabrication and
function of MEMS and NEMS such as micromirrors, nanotweezers and
nanoscale actuators \cite{13,15,16}. 
The combined action of the Casimir and
electrostatic forces can result in nonlinear dynamics,
bistable phenomena and even cause device failure by the abrupt
``pull-in'' and attachment of one surface to the other
\cite{13,15,16}. The Casimir force also changes the operation
bandwidth and tunability of MEMS. 
The actuation of MEMS using the Casimir force has been 
demonstrated \cite{16}.

The modification of the Casimir force by changing parameters
of the system other than the separation is a complicated problem 
since it requires modification of the optical properties of materials
within a relatively wide frequency region. 
The attempt to modify the Casimir force due to a coating with a
hydrogen-switchable mirror did not lead to any observed effect 
\cite{16a}.
We pioneer the
demonstration of the difference Casimir force between a gold coated sphere
and two Si samples which possess different resistivities and
charge carrier densities.
We use a high vacuum ($2\times 10^{-7}\,$Torr)
based AFM to measure the Casimir force between 
a gold coated polystyrene sphere with a diameter $2R=201.8\pm 0.6\,\mu$m
and two $4\times 7\,\mbox{mm}^2$ size Si plates placed next to each
other. The thickness of gold coating on the sphere was measured to be 
$96\pm 2\,$nm. The details of the setup were described \cite{17} in the
previous experiment with one Si plate. For this experiment
two identical polished, single crystal, $\langle 100\rangle$
orientation Si plates were chosen, 500\,$\mu$m
thick and with a resistivity $0.1-1\,\Omega\,$cm. They were $n$-type
and doped with P. The resistivity of the plates was measured using
the 4-probe technique to be 
$\tilde{\rho}\approx 0.43\,\Omega\,$cm leading to the
concentration of charge carriers 
$\tilde{n}\approx 1.2\times 10^{16}\,\mbox{cm}^{-3}$.
One of these samples was used as the first Si plate in the experiment.
The other one was subjected to thermal diffusion doping to prepare
the second, lower resistivity, plate. A phosphorous based
Spin-On-Dopant (SOD) solution (P450 commercially available from
Filmtronics Co.) was used. 
The wafers were spin-coated at
a speed of $5000\times 2\pi\,$rad/min for 0.25\,min, followed by
a pre-baking at 200\,${}^{\circ}$C for 15\,min on a hot plate. The sample
was then placed in a diffusion furnace. The diffusion was carried 
out at 1000--1050\,${}^{\circ}$C for 100 hours in a 
N${}_2$(75\%)$+$O${}_2$(25\%) atmosphere. A 49\% HF solution was
used to etch off the residual dopant after the diffusion process.
The effectiveness of the above procedure was determined using both 
a 4-probe resistivity measurement and a Hall measurement of 
a similarly doped
0.3\,$\mu$m thick single crystal Si grown epitaxially
on Si wafer. This thin equivalent sample was
homogenously doped under the above conditions \cite{17a} 
and allows a measurement of the carrier density.
The resistivity and
the carrier density were measured to be 
$\rho\approx 6.7\times 10^{-4}\,\Omega\,$cm and 
$n\approx 3.2\times 10^{20}\,\mbox{cm}^{-3}$. Both plates of 
higher and lower resistivity were subjected to a special passivation
procedure to prepare their surfaces for the force measurements. 
For this purpose nanostrip (a combination of H${}_2$O${}_2$ and
H${}_2$SO${}_4$) is used to clean the surface and 49\% HF solution
to etch SiO${}_2$ and to hydrogen terminate the surface 
\cite{17}. Finally both plates
were mounted in the AFM.

The calibration of the spring constant $k$, measurements of the
residual electrostatic potential $V_0$, deflection coefficient $m$
and separation on contact $z_0$ were done using the experimental technique
similar to what we have used in Refs.~\cite{4,17}.
 All calibration and other measurements are done in the
same high vacuum apparatus as the Casimir force measurements. 
The actual separation distance $z$ between the bottom of the gold
sphere and Si plates is given by $z=z_{\rm piezo}+mS_{\rm def}+z_0$,
where $z_{\rm piezo}$ is the distance moved by the piezo and 
$S_{\rm def}$ is the cantilever deflection signal from the
photodiodes.
First, the value of $m$ was found for the higher resistivity plate 
following the same procedure as in Ref.~\cite{17}. For this 
purpose the sphere was grounded and 29 different voltages between
--0.712 to --0.008\,V were applied to the plate through a thick
gold pad attached to plate bottom. The change in the contact
position between the sphere and the plate was used to find
$m=47.8\pm 0.2\,$nm per unit deflection signal. Then the values
of $V_0$, $km$ (which in fact is needed for force measurements)
and $z_0$ were found for the higher resistivity plate by fitting
the deflection signal $S_{\rm def}$ to the theoretical expression.
{}From the definition of the
deflection coefficient $z_d=mS_{\rm def}$ it follows
$S_{\rm def}={F_e}/{km}+S_0$.
Here the electric force between a sphere and a plate
is given by $F_e(z)=X(z)(V-V_0)^2$, where $V$ is the applied voltage and 
$X(z)$ is a known function of separation (see Refs.~\cite{3,4,9,17}
for the explicit form of $X$). 
The voltage independent offset $S_0$ represents the contribution of the 
Casimir force to the signal and was found to match the value obtained in
the independent measurement of the Casimir force. 
For the 29 different applied
voltages the measured signal $S_{\rm def}$ at every separation $z$
was plotted as a function of $V$ and fit to equation for
$S_{\rm def}$.
Note that the expression for the electric force $F_e(z)$ used in the
fit does not take into account possible influence of space-charge
layer at the surface of high-resistivity Si. According to 
Ref.~\cite{layer}, for $n$-type Si with the concentration of charge
carriers of order $10^{16}\,\mbox{cm}^{-3}$ the impact of this layer
on the electrostatic force is negligible at separations larger 
than 300--400\,nm. The use of the expression for electric force
between metal surfaces may lead to nothing more than an increased error
in the determination of $z_0$. The fit was performed within the
separation regions from 300--400\,nm to 2.5\,$\mu$m.
{}From the
fit at every $z$, the value $V_0=-0.341\pm 0.002\,$V was obtained and 
verified to be independent of $z$. The same fit results in the
values of the cantilever calibration constant multiplied
by the deflection coefficient 
$km=1.646\pm 0.004\,$pN per unit deflection signal, 
and the separation on contact $z_0=32.4\pm 1.0\,$nm.

After the calibration and related measurements for the higher resistivity 
sample are done, the Casimir force between this sample and the sphere
was measured from contact as a function of distance.
Here we report the results at $z\geq 61.19\,$nm to avoid the influence
of nonlinearities associated with the ``jump to contact'' at shorter
distances \cite{17}. 
For this purpose the sphere
was kept grounded while an appropriate compensating voltage was applied
to the plate to cancel the residual electrostatic force. The distance
between the sphere and the plate was changed continuously from large
to short separations by applying triangular voltages at 0.02\,Hz to
the piezo. The force data $F_{C,{\rm a}}^{\rm expt}(z_i)$ were collected at 
equal time intervals corresponding to equidistant points
separated by 0.17\,nm. This measurement was repeated 40 times and
the obtained forces were averaged to reduce the influence of different
random factors including thermal noise, and particular positions on
the silicon to which the sphere approaches. 
The mean values
${\bar{F}}_{C,{\rm a}}^{\rm expt}(z_i)$ of the experimental
Casimir force  data as a function of
separation are represented by dots labeled a in Fig.~1.

Next all the above calibrations and measurements were repeated for the
second, lower resistivity, Si plate. In this case 25 different dc
voltages between --0.611 to --0.008\,V were applied to the plate.
The deflection coefficient was equal to be $m=47.9\pm0.2\,$nm per unit
deflection signal. After the same fitting procedure of the measured
deflection signal, the following values of all related
parameters were obtained: $V_0=-0.337\pm 0.002\,$V,
$km=1.700\pm 0.004\,$nN per unit deflection signal, and
$z_0=32.3\pm 0.8\,$nm. The fit was performed within the separation region 
from 100--300\,nm to 2.5\,$\mu$m. Note that closer separations can be
used as the effect of the space charge layer is negligible for the
lower resistivity sample.
The values of $km$ are slightly different
in the two cases due to the changes in the cantilever level arm arising
due to minor deviations from the horizontal position in the mounting
of both samples. Next the Casimir force acting between the lower
resistivity sample and the sphere was measured from contact
after application of appropriate voltage to cancel the residual
electrostatic force. 
We report the results in a linear regime at $z\geq 60.51\,$nm.
This measurement was repeated 39 times. The
resulting mean values ${\bar{F}}_{C,{\rm b}}^{\rm expt}(z_i)$ of the
Casimir force data as a 
function of $z$ are represented in Fig.~1 by dots labeled ``b''.
As is seen from the figure, dots labeled 
``a'' and ``b'' are distinct from each
other demonstrating the effect of different charge carrier densities
in the two Si plates used.

For the quantitative characterization of the deviation between the
two measurements, we calculate the random errors using the 
procedure outlined in Ref.~\cite{17} based on the Student's $t$
distribution. For the sample of higher resistivity (measurement ``a'')
the random error at 95\% confidence is equal to 8\,pN at $z=61.19\,$nm,
decreases to 6\,pN at $z=70\,$nm and becomes equal to 4\,pN at
$z\geq 80\,$nm. The measurement ``b'' for the sample of lower resistivity
is slightly more noisy. Here the random error at 95\% confidence changes
from 11\,pN  at $z=60.51\,$nm, 7\,pN at $z=70\,$nm  to 5\,pN at
$z\geq 80\,$nm. The systematic error at 95\% confidence is 
equal to only 1.2\,pN for both measurements 
(see Ref.~\cite{17} for details). Using the 
statistical criterion in Ref.~\cite{20}, we conclude that the total
experimental errors $\Delta F_{\rm a,b}$ determined at 95\% confidence
are equal to the random ones in each measurement. From Fig.~1 it is seen
that the deviation between the two sets of data is 
larger than the total experimental error in the separation region from
61.19 to 120\,nm.

Now we compare the force-distance relation measured for the two Si
samples with the theory. At $z<150\,$nm, where the differences 
between the two measurements are most pronounced, the magnitudes of the
predicted thermal corrections are negligible \cite{17}. 
At larger $z$ the relative contribution from thermal corrections
is much less than the relative error of force measurements.
Then the force 
between the sphere and one of the plates ($\alpha=\mbox{a}$ for higher
and $\alpha=\mbox{b}$ for lower resistivity Si) is given by the Lifshitz
formula at zero temperature adapted for the configuration of a sphere
above a plate \cite{9,22}
\begin{equation}
F_{\alpha}(z)=\frac{\hbar R}{2\pi}\int_{0}^{\infty}k_{\bot}dk_{\bot}
\int_{0}^{\infty}d\xi\,
\sum\limits_{\kappa=\|,\bot}\ln\left[1-
r_{\kappa}^{(1)}(\xi,k_{\bot})r_{\kappa,\alpha}^{(2)}(\xi,k_{\bot})
e^{-2zq}\right].
\label{eq2}
\end{equation}
\noindent
The reflection coefficients $r_{\|,\bot}^{(1)}$ for gold and
$r_{\|,\bot;\alpha}^{(2)}$ for the two types of Si are expressed in the usual 
way \cite{17} through the dielectric permittivities of gold
$\varepsilon^{(1)}(i\xi)$ and of Si $\varepsilon_{\alpha}^{(2)}(i\xi)$
along the imaginary frequency axis 
($q^2\equiv k_{\bot}^2+\xi^2/c^2$).
The permittivities $\varepsilon^{(1)}(i\xi)$ and 
$\varepsilon_{\rm a}^{(2)}(i\xi)$ are computed by means of the dispersion
relation using the tabulated optical data for the complex
index of refraction \cite{23}. The results are shown in Fig.~2
with the solid line and dashed line labeled ``a'', respectively \cite{24}.
The permittivity of lower resistivity Si is found from \cite{23}
$\varepsilon_{\rm b}^{(2)}(i\xi)={{\varepsilon}}_{\rm a}^{(2)}(i\xi)+
{\omega_p^2}/{[\xi\left(\xi+\gamma\right)]}$,
where $\omega_p$ and $\gamma$ are
the plasma frequency and relaxation parameter of lower resistivity Si.
Their values were found from
$\omega_p=e\sqrt{n}/\sqrt{\varepsilon_0m^{\ast}}
\approx 2.0\times 10^{15}\,$rad/s and
$\gamma=\varepsilon_0\rho\omega_p^2\approx 2.4\times 10^{14}\,$rad/s,
where $m^{\ast}=0.26m_e$ is the electron effective mass and
$\varepsilon_0$ is the permittivity of vacuum.
The permittivity $\varepsilon_{\rm b}^{(2)}$ is shown in Fig.~2
by the dashed line labeled ``b''. Then the forces 
$F_{\rm a}(z)$ and $F_{\rm b}(z)$ were calculated 
at all separations using Eq.~(\ref{eq2}).

The obtained results were corrected for the presence of surface 
roughness. To do this, the topographies of the sphere and both Si
samples were investigated with an AFM. Then the scan data were used
to additively compute the Casimir forces
$F_{\rm a}^{\rm theor}(z)$ and $F_{\rm b}^{\rm theor}(z)$ starting from 
$F_{\rm a}(z)$ and $F_{\rm b}(z)$ and varying separations in accordance
with the roughness profiles. The details of this procedure
and justification of the additive approach for the experimental
situation can be found in Refs.~\cite{4,8,17}.
For both plates the contribution of the roughness to the Casimir force
was equal. 
It changes from 3.6\% of the total force at $z=60\,$nm to 2.7, 1.4 and
0.65\% at separations 70, 100 and 150\,nm, respectively.
Surface distortions on single crystal Si are
very low and practically do not contribute to the roughness correction.

The errors in the computation of the Casimir force between the gold coated 
sphere and Si plate are analyzed in Ref.~\cite{17}.
At the shortest $z$ they are mostly determined by the error
$\Delta z=1.0\,$nm (the plate ``a'') and 0.8\,nm (the plate ``b'') 
in the measurement of separations $z_i$ with which
the theoretical values of the Casimir force are calculated for
the comparison with the experiment. 
A 0.5\% error due to the variation of optical parameters \cite{4}
is also included.
At $z=60\,$nm the total
theoretical error at 95\% confidence is equal to 19.6\,pN
(4.9\% of the force) for the plate ``a'' and to 17.2\,pN (4.0\% of the
force) for the plate ``b''. 
It decreases to 11\,pN (4.2\% of the force) for the plate ``a'' and to
9.6\,pN (3.4\% of the force) for the plate ``b''
at $z=70\,$nm. The total theoretical error becomes less
than the total experimental error at $z>90$ and 85\,nm for the plates
``a'' and ``b'', respectively.

The total theoretical error was combined with the total experimental 
error at 95\% confidence using the statistical rule in Ref.~\cite{20}
(see also Refs.~\cite{8,17}) to find the error $\Xi(z)$ of the difference
between theoretical and experimental forces. The obtained confidence
interval $[-\Xi(z),\Xi(z)]$ as a function of separation is shown
in Fig.~3 as solid lines. The differences 
 $F_{\rm a}^{\rm theor}(z)-F_{\rm a}^{\rm expt}(z)$ versus separation
for the experiment with the higher resistivity Si are plotted in Fig.~3a
as dots. Similarly, the differences 
 $F_{\rm b}^{\rm theor}(z)-F_{\rm b}^{\rm expt}(z)$
for the lower resistivity Si are shown as dots in Fig.~3b.
As is seen from Fig.~3, both measurements are consistent with
theories using the dielectric permittivity 
$\varepsilon_{\rm a}^{(2)}(i\xi)$ (Fig.~3a) and
$\varepsilon_{\rm b}^{(2)}(i\xi)$ (Fig.~3b).

To illustrate the effect of modification of the Casimir force through 
the change of carrier density, in Fig.~4 we plot as dots the difference 
of the measured Casimir forces for the plates of lower and higher
resistivities,  $F_{\rm b}^{\rm expt}(z)-F_{\rm a}^{\rm expt}(z)$,
versus separation. In the same figure, the difference in the respective
theoretically computed Casimir forces,
$F_{\rm b}^{\rm theor}(z)-F_{\rm a}^{\rm theor}(z)$ is shown as the
solid line. As is seen in Fig.~4, the experimental and theoretical
difference Casimir forces as functions of $z$ are in good
agreement. It can be easily shown that the magnitude of the mean difference
of the measured Casimir forces exceeds the 
experimental error of force difference 
within the separations from 70 to 100\,nm.

To conclude, we have measured the Casimir force between a gold coated 
sphere and two Si plates of higher and lower resistivity 
differing by several orders of
magnitude. Each measurement was compared with theoretical results
using the Lifshitz theory with different dielectric permittivities and
found to be consistent with it. The difference of the measured forces
for the two resistivities is in good agreement with the 
corresponding difference of the theoretical results.
It takes a magnitude of about 17\,pN at 
$z=70\,$nm and decreases with increase of separation.
The performed experiment demonstrates the possibility to modify
the Casimir force by changing the carrier density of semiconductor
materials which may find applications in the design, fabrication and
function of MEMS and NEMS. 

%%%%%%%%%%%%%%%%%%%%%%%%%%%%%%%%%%%%%%%%%
%\section*{Acknowledgment}
This work was supported by the NSF Grant PHY0355092 and
DOE grant DE-FG02-04ER46131.

%%%%%%%%%%%%%%%%%%%%%%%%%%%%%%%%%%%%%%%%%%

%%%%%%%%%%%%%%%%%%%%%%%%
%\end{document}
%%%%%%%%%%%%%%%%%%%%%
%%%%%%%%%%%%%%%%%%%%%%%%%%%%%%%%%%
%%%__FIGURES__%%%%%%%%%%
%%%%%%%%%%%%%%
\begin{figure}
\vspace*{-7cm}
\centerline{
\includegraphics{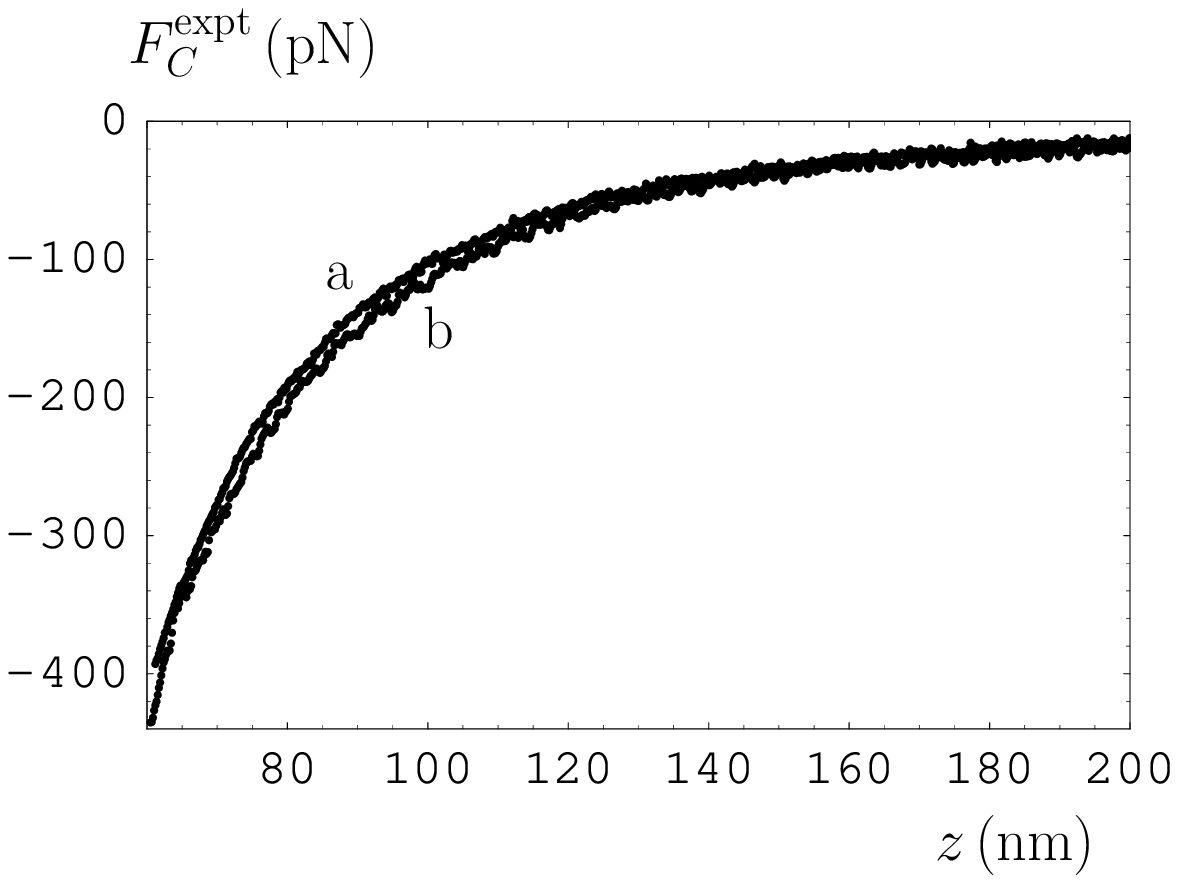}
}
\vspace*{-8cm}
\caption{The data of the mean measured Casimir force as a function 
of separation
between a gold coated sphere and two Si plates of (a) higher and
(b) lower resistivities are shown as dots.}
\end{figure}
%%%%
\begin{figure}
\vspace*{-7cm}
\centerline{
\includegraphics{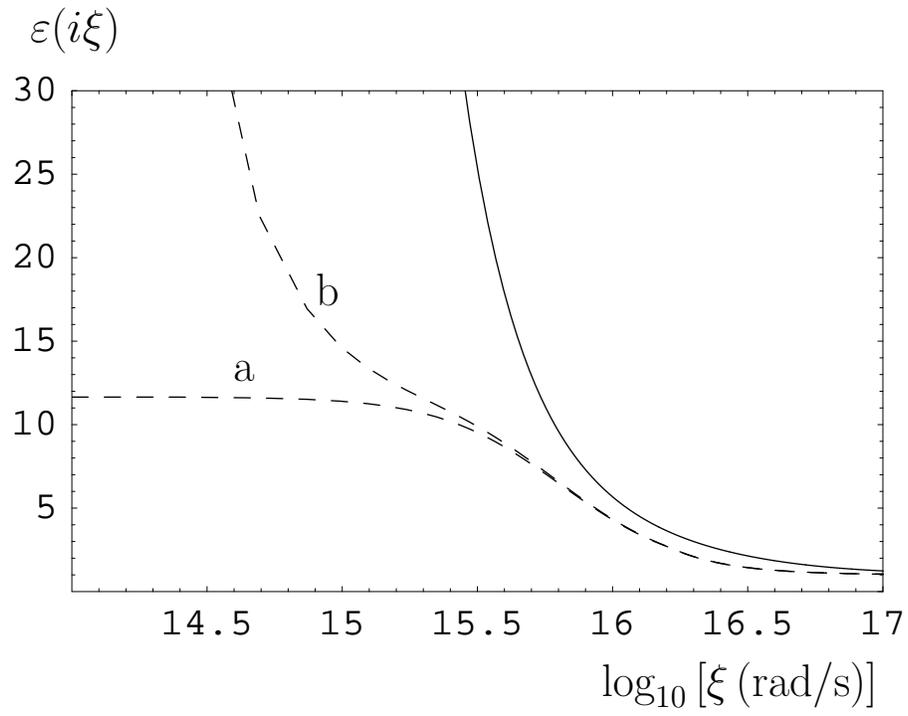}
}
\vspace*{-8cm}
\caption{Dielectric permittivities along the imaginary frequency axis for
gold (solid line) and for Si (dashed lines ``a'' and ``b'' are for higher
and lower resistivity Si, respectively).}
\end{figure}
%%%%
\begin{figure}
\vspace*{-3cm}
\centerline{
\includegraphics{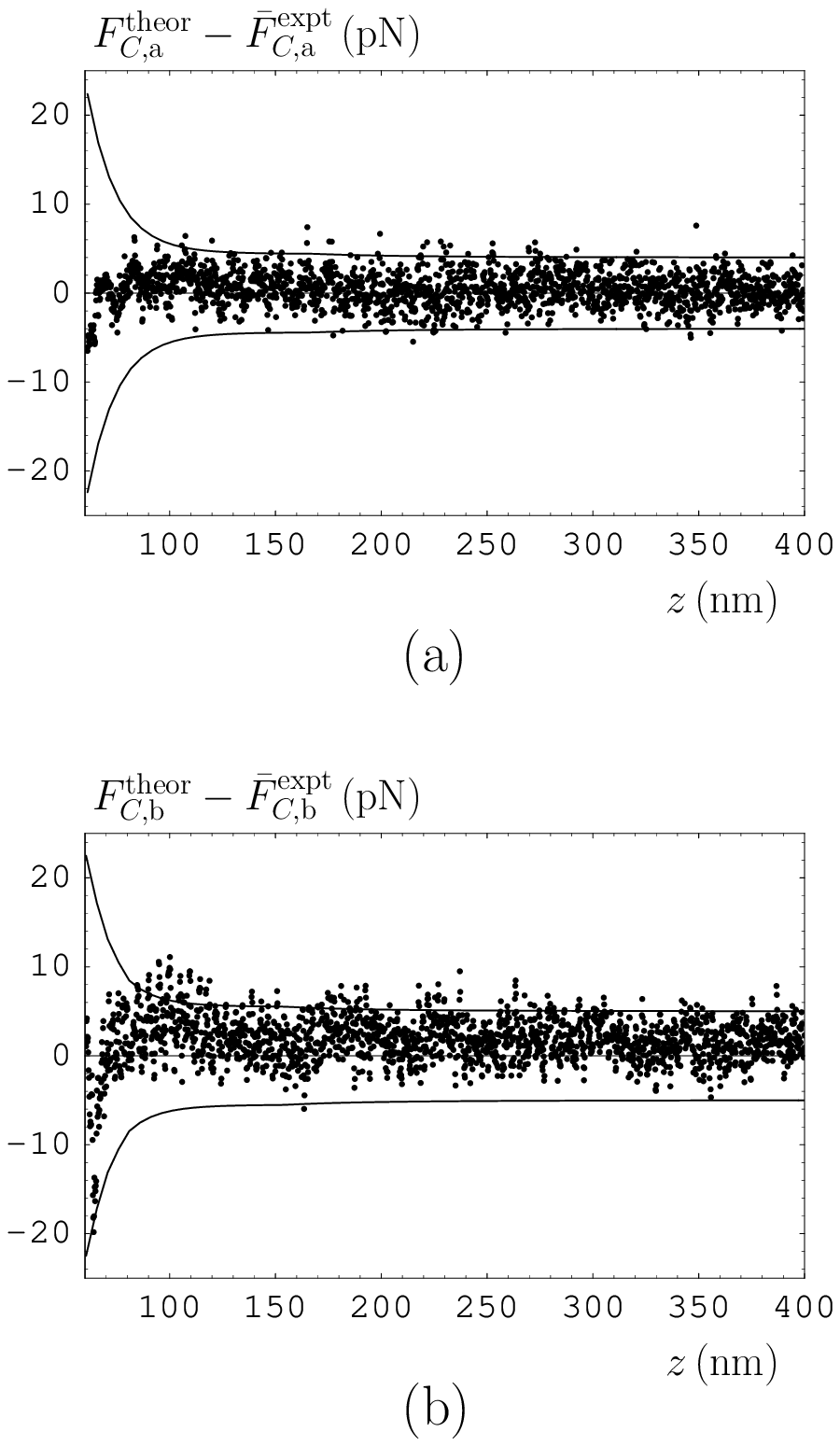}
}
\vspace*{-8cm}
\caption{Differences of the theoretical and mean experimental 
Casimir forces versus separation. Forces are computed and measured for
(a) higher and (b) lower resistivity Si. Solid lines indicate
95\% confidence intervals.} 
\end{figure}
%%%%
\begin{figure}
\vspace*{-7cm}
\centerline{
\includegraphics{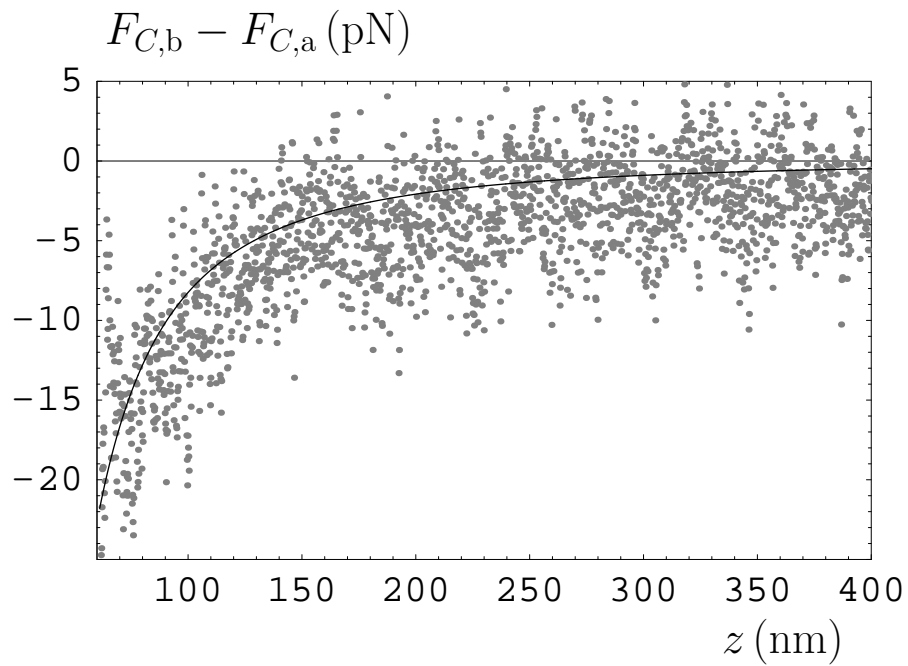}
}
\vspace*{-8cm}
\caption{The differences of the mean measured Casimir forces of
the lower and higher resistivity Si samples versus separation 
are shown as dots.
The corresponding theoretically calculated 
differences are shown by the solid line.} 
\end{figure}
%%%%
\end{document}